# An Asynchronous Automata Approach to the Semantics of Temporal Logic

**Serban E. Vlad**
Oradea City Hall
E-mail: serbanvlad@excite.com

**Abstract**
The paper presents the differential equations that characterize an asynchronous automaton and gives their solution $x: \mathbf{R} \to \{0,1\}^n$. Remarks are made on the connection between the continuous time and the discrete time of the approach. The continuos and discrete time, linear and branching temporal logics have the semantics depending on $x$ and their formulas give the properties of the automaton.

## 1. Preliminaries

1.1 We note $\mathbf{B}_2 = \{0,1\}$ together with the *complement* '$\bar{\ }$', the *product* '$\cdot$', the *modulo 2 sum* '$\oplus$' etc, the *order* $0 \leq 1$ and the *discrete topology*.

1.2 The function $w: \mathbf{R} \to \mathbf{B}_2$ is called *realizable*, if it may be put under the form:
$$w(t) = w(z_0) \cdot \chi_{[z_0, z_1)}(t) \oplus w(z_1) \cdot \chi_{[z_1, z_2)}(t) \oplus ..., t \in \mathbf{R} \qquad (1)$$
where the real family $\{z_k \mid k \in \mathbf{N}\}$ is *strictly increasing non-negative locally finite* SINLF, i.e. $0 = z_0 < z_1 < z_2 < ...$ and
$$\forall a < b, (a,b) \wedge \{z_k \mid k \in \mathbf{N}\} \text{ is finite}$$
are satisfied and $\chi_{[z_k, z_{k+1})}: \mathbf{R} \to \mathbf{B}_2$ is the characteristic function of the interval $[z_k, z_{k+1})$.

1.3 We note with $Real$ the ring of the realizable functions and with $Real^{(n)}$ the linear space $\{(w_1(t), ..., w_n(t)) \mid w_1, ..., w_n \in Real\}$.

1.4 For an arbitrary function $w(t)$, the *left limit* $w(t-0)$ and the *left derivative* $D^- w(t)$ functions are defined in the following manner: for any $t$
$$\exists t' < t, \forall \xi \in (t', t), w(\xi) = w(t-0) \qquad (1)$$
$$D^- w(t) = w(t-0) \oplus w(t) \qquad (2)$$

and similar definitions hold for the right duals $w(t+0)$ and $D^+w(t)$.

1.5 If $w \in Real$, then all the previous four functions exist. We shall consider that the realizable functions are models for the electrical signals of the asynchronous circuits.

## 2. Delays

2.1 The delays from the asynchronous circuits, occuring on gates and on wires, are by definition characterized by the next equations:

i) the *ideal* delays

$$x(t) = f(u_1(t-\tau),...,u_m(t-\tau)) \cdot \chi_{[\tau,\infty)}(t) \tag{1}$$

ii) the *inertial* delays

$$D^- x(t) = (x(t-0) \oplus \hat{f}(u_1(t-0),...,u_m(t-0),t-0)) \cdot$$
$$\cdot \overline{\bigcup_{\xi \in (t-\tau,t)} D^- \hat{f}(u_1(\xi),...,u_m(\xi),\xi)} \tag{2}$$

where $\bigcup$ is the maximum of a function on a set; $u_1,...,u_m \in Real$ are the models of the input electrical signals; $x \in Real$ is the model of the delayed state (or output) electrical signal; $f : \boldsymbol{B}_2^m \to \boldsymbol{B}_2$ is the Boolean function that is implemented and $\hat{f}(u_1(t),...,u_m(t),t)$ is defined in the following manner:

$$\hat{f}(u_1(t),...,u_m(t),t) = f(u_1(t),...,u_m(t)) \cdot \chi_{[0,\infty)}(t) \tag{3}$$

$\tau > 0$ is the delay parameter.

2.2 The delay parameters $\tau_1,...,\tau_n$ are not known. They depend (at least) on the technology, on the temperature and on the sense of the switch, from 0 to 1, respectively from 1 to 0. The technologist indicates lower bounds and upper bounds for them; if such bounds are available, $\tau_1,...,\tau_n$ are called *bounded* and if not, they are called *unbounded*.

2.3 Our present purpose is to interpret the inertial delays described by 2.1 (2). We shall suppose for this the existence of $c \in \boldsymbol{B}_2$ and of $\tau \leq t_1 < t_2$ so that

$$\forall t \in [t_1 - \tau, t_1) \vee [t_2, \infty), f(u_1(t),...,u_m(t)) = c \tag{1}$$

$$\forall t \in [t_1, t_2), f(u_1(t),...,u_m(t)) = \overline{c} \tag{2}$$

i.e. for $t \geq t_1 - \tau$, $f(u_1(t),...,u_m(t))$ has a constant value equal to $c$, with the exception of the interval $[t_1, t_2)$ when a perturbation occurs and the function changes its value to $\overline{c}$.

We infer, after a few computations:

$$\bigcup_{\xi \in (t_1-\tau,t_1)} D^- f(u_1(\xi),...,u_m(\xi)) = \bigcup_{\xi \in (t_1-\tau,t_1)} D^- c(\xi) = 0 \tag{3}$$

$$x(t_1) = c = f(u_1(t_1-\tau),...,u_m(t_1-\tau)) \tag{4}$$

There exist two possibilities.

Case I $t_2 - t_1 < \tau$ gives

$$t \geq t_1, x(t) = c \tag{5}$$

meaning that the perturbation is eliminated (it is filtered).

Case II $t_2 - t_1 \geq \tau$ implies

$$\bigcup_{\xi \in (t_1, t_1+\tau)} D^- f(u_1(\xi),...,u_m(\xi)) = \bigcup_{\xi \in (t_1, t_1+\tau)} D^- \overline{c}(\xi) = 0 \tag{6}$$

$$x(t_1 + \tau) = \overline{c} = f(u_1(t_1),...,u_m(t_1)) \tag{7}$$

$$\bigcup_{\xi \in (t_2, t_2+\tau)} D^- f(u_1(\xi),...,u_m(\xi)) = \bigcup_{\xi \in (t_2, t_2+\tau)} D^- c(\xi) = 0 \tag{8}$$

$$x(t_2 + \tau) = c = f(u_1(t_2),...,u_m(t_2)) \tag{9}$$

and finally

$$x(t) = \begin{cases} c, t \in [t_1, t_1 + \tau) \vee [t_2 + \tau, \infty) \\ \overline{c}, t \in [t_1 + \tau, t_2 + \tau) \end{cases} \tag{10}$$

i.e. the perturbation is delayed with $\tau$ time units.

The equations (3), (6), (8) represent conditions of balance for the system and this balance implies the validity of (4), (7), (9).

**3. Asynchronous Automata**

3.1 We call *asynchronous automaton*, or *asynchronous system*, a mathematical object $\Sigma$ given by the following data: $\mathbf{R}$ is the time set and $0 \in \mathbf{R}$ is the *initial time*; $\mathbf{B}_2^n$ is the state space, where $n \geq 2$; $x \in Real^{(n)}$ is the *state* and $x^0 \in \mathbf{B}_2^n$ is called the *initial state*; $\mathbf{B}_2^m$ is the input space and $u \in Real^{(m)}$ is the *input*; the function $f : \mathbf{B}_2^n \times \mathbf{B}_2^m \to \mathbf{B}_2^n$, is called the *generator function*; the numbers $\tau_i > 0, i = \overline{1,n}$ are called the *delays*; the next equations are called the *equations of the asynchronous automata*, shortly EAA:

$$x_i(t) = f_i(x(t-\tau_i), u(t-\tau_i)) \cdot \chi_{[\tau_i, \infty)}(t) \oplus x_i^0 \cdot \chi_{[0,\tau_i)}(t), i = \overline{1, n_1} \tag{1}$$

$$D^- x_i(t) = (x_i(t-0) \oplus f_i(x(t-0), u(t-0))) \cdot \tag{2}$$

$$\cdot \bigcup_{\xi \in (t-\tau_i, t)} D^- f_i(x(\xi), u(\xi)) \cdot \chi_{[\tau_i, \infty)}(t) \oplus x_i^0 \cdot \chi_{\{0\}}(t), i = \overline{n_1 + 1, n}$$

where $t \in \mathbf{R}$ and $1 \leq n_1 < n$; $x^0$ and $u$ are given, $\tau_1,...,\tau_n$ are parameters and $x$ is the unknown. The coordinates $i \in \{1,...,n_1\}$ are called *ideal* and the coordinates $i \in \{n_1 + 1,...,n\}$ are called *inertial*.

3.2 **Remark**, special cases for EAA

a) the case when we have only ideal coordinates, respectively only inertial coordinates may be put under the form 3.1 (1), (2) by adding a null coordinate; this is possible because the null function $x_i(t) = 0$ is a solution for any of 3.1 (1), (2) when $x_i^0 = 0$ and $f_i = 0$, thus EAA with $1 \leq n_1 < n$ is not restrictive.

Conversely, if the automaton has no ideal coordinates, respectively no inertial coordinates, then the missing coordinates don't have to be included in EAA as null coordinates.

b) the case of the *autonomous automata*, when there does not exist an input; we suppose that $u = 0 \in \boldsymbol{B}_2^m$

c) the case of the *trivial automata*, when there does not exist a state; we have that $x = 0 \in \boldsymbol{B}_2^n$.

**3.3 Remark** There exist other similar ways of writing EAA, that bring nothing essentially new, see 2.1. For example, it is possible to initialize the coordinates $x_1, ..., x_n$ not starting with the time instant $t_0 = 0$ but ending with the time instant $t_1$, see 2.3 (1) and the consequence 2.3 (4).

**3.4 Remark** There exists a right dual, anticipative version of EAA.

**3.5 Theorem** We suppose that in EAA

$$u(t) = u^0 \cdot \chi_{[v_0, v_1)}(t) \oplus u^1 \cdot \chi_{[v_1, v_2)}(t) \oplus ... \qquad (1)$$

where $u^k \in \boldsymbol{B}_2^m, k \in \boldsymbol{N}$ and $0 = v_0 < v_1 < v_2 < ...$ is SINLF (see 1.2). We note the elements of the set $\{v_k + p_1 \cdot \tau_1 + ... + p_n \cdot \tau_n \mid k, p_1, ..., p_n \in \boldsymbol{N}\}$ under the form of the SINLF family $0 = t_0 < t_1 < t_2 < ...$ Then EAA has a unique solution

$$x(t) = x^0 \cdot \chi_{[t_0, t_1)}(t) \oplus x^1 \cdot \chi_{[t_1, t_2)}(t) \oplus ... \qquad (2)$$

where $x^k \in \boldsymbol{B}_2^n$ satisfy for all $k \in \boldsymbol{N}$:

$$x_i^{k+1} = \begin{cases} x_i^0, \text{if } t_{k+1} < \tau_i \\ f_i(x(t_{k+1} - \tau_i), u(t_{k+1} - \tau_i)), \text{if } t_{k+1} \geq \tau_i \end{cases} \qquad (3)$$

$\forall i \in \{1, ..., n_1\}$ and

$$x_i^{k+1} = \begin{cases} x_i^0 & , \text{if } t_{k+1} < \tau_i \\ f_i(x(t_k), u(t_k)) & , \text{if } t_{k+1} \geq \tau_i \text{ and} \\ \quad \forall t_s, t_q \in [t_{k+1} - \tau_i, t_{k+1}), f_i(x(t_s), u(t_s)) = f_i(x(t_q), u(t_q)) \\ x_i^k & , \text{if } t_{k+1} \geq \tau_i \text{ and} \\ \quad \exists t_s, t_q \in [t_{k+1} - \tau_i, t_{k+1}), f_i(x(t_s), u(t_s)) \neq f_i(x(t_q), u(t_q)) \end{cases} \qquad (4)$$

$\forall i \in \{n_1 + 1, ..., n\}$.

## 4. Continuous Time, or Discrete Time ?

4.1 Our answer to the previous question is: both, because they coexist for the realizable functions. In fact, EAA have been written in continuous time and their solution was given at 3.5 (3), (4) in discrete time.

4.2 The fact that $\tau_1,...,\tau_n$ are not known, thus the sampling moments $\{t_k \mid k \in N\}$ are not known, gives another perspective on the discrete time. On the other hand, we observe that even if $\{z_k \mid k \in N\}$ are known in 1.2 (1), they are not unique for some arbitrary realizable function.

4.3 In [6], Alfaro and Manna put the problem of discrete reasoning in continuous time. They show that if a formula of the continuous temporal logic has the property of finite variability FV, then its validity in the discrete semantics implies the one in the continuous semantics. The condition FV is similar to the local finiteness condition that we have put: $\forall a < b, \ (a,b) \wedge \{z_k \mid k \in N\}$ is finite.

## 5. Propositional Linear Time Temporal Logic

5.1 The Boolean variables $x_1,...,x_n \in \boldsymbol{B}_2$ are also called *atomic propositions* of the classical logic of the propositions CLP. The Boolean functions $h: \boldsymbol{B}_2^n \to \boldsymbol{B}_2$ are also called *formulas* of CLP[1].

5.2 The semantic approach of CLP answers the question: in the interpretation $I$ that gives the variable $x = (x_1,...,x_n)$ the constant value $x^0 = (x_1^0,...,x_n^0) \in \boldsymbol{B}_2^n$ do we have $h(x^0) = 1$? If so, we say that $h$ *is satisfied in* $I$ or that it *holds in* $x^0$ and we note this fact with $x^0 \models h$. For $h = 1$ (the constant function), we say that it is a *tautology* and we note this fact by $\models h$.

5.3 In the Linear Time Temporal Logic LTL, the *atomic propositions* are the functions $x_1,...,x_n \in Real$ and some of the *formulas*, that are induced by the laws of $\boldsymbol{B}_2$, are obtained by associating to the functions $\boldsymbol{B}_2^n \to \boldsymbol{B}_2$ respectively functions $Real^{(n)} \to Real$. For example, to $x_1 \cdot x_2, x_1 \oplus x_2$ where $x_1, x_2 \in \boldsymbol{B}_2$ we associate the functions $(x_1 \cdot x_2)(t) = x_1(t) \cdot x_2(t)$, $(x_1 \oplus x_2)(t) = x_1(t) \oplus x_2(t)$ where $x_1, x_2 \in Real$.

5.4 It is interesting the semantic approach of LTL answering the question: in the interpretation that gives the argument $x \in Real^{(n)}$ of $h$ the constant value, noted with the same symbol, representing the solution of EAA and fixes the time to $t \geq 0$, do we have that $h(x(t)) = 1$? If so, we say that $h$ *is satisfied at* $t$, or that it *holds at* $t$ and we note this fact with $t \models h$. In the discrete version, the equation $h(x(t_k)) = 1$ is noted with $k \models h$. In both versions, instead of $0 \models h$ we write $\models h$.

---

[1] This definition identifies the logically equivalent formulas.

5.5 **Remark** The semantics that we use here is called *floating* and it differs from the *anchored* semantics by the fact that the latter refers only to statements of the type $\models h$.

5.6 LTL has in its discrete version the unary temporal connector $X$, called *Next* (noted sometimes with $O$) with a semantics defined like this: the equation $h(x(t_{k+1})) = 1$ is noted $k \models Xh$. Alfaro and Manna just mention the fact that in their theory $X$ is missing. The continuous semantics of this connector is rather given by the equation $h(x(t-0)) = 1$, noted $t \models h^-$, then by $h(x(t+0)) = 1$, noted $t \models h^+$, as it would seem to be normal; the realizable functions are right continuous and the connector $h^+ = h$ is of null effect in the non-anticipative reasoning.

5.7 LTL has also the binary temporal connector *Until* $U$, which is present in both continuous respectively discrete version, for example:
$$\bigcup_{t' \geq t} g(x(t')) \cdot \bigcap_{\xi \in [t,t')} h(x(\xi)) = 1 \text{ is noted with } t \models h\,U\,g$$
where $\bigcap$ is the minimum of a function on a set; if this set is empty, for $t' = t$, then by definition the minimum is taken to be equal with $1$.

5.8 $U$ gives the possibility of defining the unary connectors *Always*, *Henceforth*, or *Necessity* $G$, respectively *Sometimes*, *Eventually*, or *Possibility* $F$. If in 5.7 $h = 1$:
$$\bigcup_{t' \geq t} g(x(t')) = 1 \text{ is noted with } t \models Fg$$

5.9 Alfaro and Manna mention in the syntax of their temporal logic the *age function* $\Gamma$: "*for a formula $h$, at any point in time, the term $\Gamma(h)$ denotes for how long in the past $h$ has been continuously true*". Let us remark that such an idea occurs in EAA in the term $\bigcup_{\xi \in (t-\tau_i, t)} D^- f_i(x(\xi), u(\xi))$: if $f_i(x(\xi), u(\xi))$ is *constant* on the interval $(t - \tau_i, t)$, its derivative is null on this interval, the reunion is null also and its complement is unitary; this is a necessary condition (of inertiality) to have $D^- x_i(t) = 1$. Thus the asynchronous automata make use of age functions $\Gamma_i$ which replace in their definition 'continuously true' with 'continuously constant' and limit coordinatewise the memory of the automaton to $\tau_i$ time units.

## 6. Propositional Branching Time Temporal Logic

6.1 By definition, the trajectory $x \in Real^{(n)}$ of an asynchronous automaton $\Sigma$ may take after $x^0$ any of the values:

$t = t_1$:
$(f_1(x^0, u^0), x_2^0, ..., x_n^0), ..., (x_1^0, ..., f_i(x^0, u^0), ..., x_n^0), ..., (f_1(x^0, u^0), ..., f_n(x^0, u^0))$

$t = t_2$: $\quad (f_1(f_1(x^0, u^0), x_2^0, ..., x_n^0), u(t_1)), x_2^0, ..., x_n^0), ...$

$..., (f_1(x^0, u^0), ..., f_i((f_1(x^0, u^0), x_2^0, ..., x_n^0), u(t_1)), ..., x_n^0), ...$

$..., (f_1((f_1(x^0, u^0), ..., f_n(x^0, u^0)), u(t_1)), ..., f_n((f_1(x^0, u^0), ..., f_n(x^0, u^0)), u(t_1)))$

...

By *branching time* it is understood the common picture of all these possibilities.

6.2 **Remark** We can similarly suppose that unlike the previous paragraphs, $\tau_1, ..., \tau_n$ are parametric piecewise constant $\boldsymbol{R} \to (0, \infty)$ functions. Then $x$ is the solution of an equation EAA' differing from EAA by this generalization and where $\{t_k \mid k \in \boldsymbol{N}\}$ is the SINLF family that counts the elements of the set

$$\{v_k + p_1 \cdot \tau_1(t) + ... + p_n \cdot \tau_n(t) \mid k, p_1, ..., p_n \in \boldsymbol{N}, t \in \boldsymbol{R}\}$$

When $\tau_1, ..., \tau_n$ vary, a family of trajectories results.

6.3 A *path* is one of the possible trajectories of the automaton, when $\tau_1, ..., \tau_n$ vary as parameters, i.e. a function $x(t)$ for the continuous time and a sequence $(x^0, x^1, x^2, ...)$ for the discrete time. Let *Path* be the set of all the paths of the automaton $\Sigma$.

6.4 The branching temporal logics, for example $BT, BT^+, UB, UB^+, CTL, CTL^+, CTL^*$ [7], $\forall CTL, \exists CTL, \forall CTL^*, \exists CTL^*$ [5] have common features. In their syntax appear like at the linear time temporal logics the Boolean connectors, as well as the temporal connectors $F, G, X, U$ regarded to be *state quantifiers*. In addition we have the *path quantifiers* $A$, *for all paths* and $E$, *for some path*.

6.5 The semantics of $A$ and $E$ is naturally defined. For example:
$$\bigcap_{x \in Path} h(x(t)) = 1 \text{ is noted with } t \models Ah$$

**Bibliography**

1. Serban E. Vlad, Pseudoboolean Field Lines, Proceedings of the XX-th National Conference of Geometry and Topology, Timisoara, October 5-7, 1989

2. Serban E. Vlad, Asynchronous Automata: Differential Equations, the Technical Condition of Good Running and the Stability, in Cybernetics and Systems Research vol.1, Robert Trappl Editor, World Scientific, 1992

3. Luciano Lavagno, Synthesis and Testing of Bounded Wire Delay Asynchronous Circuits from Signal Transition Graphs, Ph.D. Thesis, University of California at Berkeley, 1992

4. Moshe Vardi, An Automata Theoretic Approach to Linear Temporal Logic, Banff'94

5. O. Kupferman, M.Y.Vardi, Relating Linear and Branching Model Checking, Chapman & Hall, IFIP, 1996

6. Luca de Alfaro, Zohar Manna, Verification in Continuous Time by Discrete Reasoning, Proceedings of the 4-th International Conference on Algebraic Methodology and Software Technology AMAST'95, Lecture Notes in Computer Science 936, pages 292-306, Springer Verlag, 1995.

7. Aarti Gupta, Formal Hardware Verification Methods: A Survey, CMU-CS-91-193, Technical report, 1991

8. Mihaela Malita, Mircea Malita, Bazele inteligentei artificiale (The Basic Elements of Computer Science), ed. Tehnica, Bucuresti, 1987